\documentclass[12pt,reqno]{amsart}
\usepackage{epsfig}
\usepackage{graphicx}
\begin{document}
\baselineskip=0.8cm 
\theoremstyle{plain}
\newtheorem{thm}{Theorem}[section]
\newtheorem{lem}{Lemma}[section]
\newtheorem{prop}{Proposition}[section]
\newtheorem{coll}{Conclusion}
\theoremstyle{remark}
\newtheorem{rem}{Remark}[section]

\title{Coupled KdV equations derived \\ from atmospherical dynamics}
\author{S. Y. Lou$^{1,2}$, Bin Tong$^1$, Heng-chun Hu$^{1,3}$,\\
Xiao-yan Tang$^1$} \dedicatory{$^1$Department of Physics, Shanghai
Jiao Tong
University, Shanghai, 200030, China\\
$^2$Center of Nonlinear Science, Ningbo University, Ningbo,
315211, China\\
$^3$Department of Mathematics, Shanghai Science and Technology
University, Shanghai, China}
\begin{abstract}
Some types of coupled Korteweg de-Vries (KdV) equations are
derived from an atmospheric dynamical system. In the derivation
procedure, an unreasonable $y$-average trick (which is usually
adopted in literature) is removed. The derived models are
classified via Painlev\'e test. Three types of $\tau$-function
solutions and multiple soliton solutions of the models are
explicitly given by means of the exact solutions of the usual KdV
equation. It is also interesting that for a non-Painlev\'e
integrable coupled KdV system there may be multiple soliton
solutions.
\end{abstract}
 \maketitle

\section{Introduction}
The single component KdV equation has been widely used in various
natural science fields especially in almost all branches of
physics. For instance \cite{Kivshar}, the KdV equation describes,
in a general form, competition between weak nonlinearity and weak
dispersion, while the nonlinear Schr\"odinger (NLS) equation
describes the same competition for envelope waves (see, for
example, the Introduction in \cite{1980}). Some other integrable
equations such as the sine-Gordon (SG) equation, the
Kadomtsev-Petviashvily equation, the so-called three- and
four-wave systems are universal as well.

Some kinds of the coupled KdV equations had also been introduced
in literature such as one describing two resonantly interacting
normal modes of internal-gravity-wave motion in a shallow
stratified liquid \cite{CKdV}. In principle, many of other coupled
KdV equations are introduced mathematically because of their
integrability \cite{CK}.

In section II of this paper, we derive some new types of coupled
KdV equation systems with some arbitrary parameters from a
two-layer fluid model which is used to describe the atmospheric
phenomena by using long wave approximation.

Whence the coupled KdV systems are derived, an important problem
is how to solve them. To get more exact solutions, one hopes to
figure out the integrable ones. In section III of the paper, we
 use the well known Painlev\'e test classification to find out
the Painlev\'e integrable ones for special types of selections of
the parameters.

For some special types of coupled KdV systems, one can find some
types of exact solutions by modifying the solutions of the usual
KdV equation. Some concrete solution examples, especially, the
$\tau$-function solutions and the multiple soliton solutions are
given in section V. The last section contains a short summary and
discussion.

\section{Coupled KdV equations derived from atmospheric dynamics}
It is known that various integrable models can be derived from
fluid dynamics. Similarly, in this section we use a two layer
fluid model,
\begin{eqnarray}
&&q_{1t}+J\{\psi_{1},\
q_1\}+\beta\psi_{1x}=0,\label{eq1}\\
&&q_{2t}+J\{\psi_{2},\ q_2\}+\beta\psi_{2x}=0,\label{eq2}
\end{eqnarray}
where
\begin{eqnarray}
&&q_1=\psi_{1xx}+\psi_{1yy}+F(\psi_2-\psi_1),\label{q1}\\
&&q_2=\psi_{2xx}+\psi_{2yy}+F(\psi_1-\psi_2),\label{q2}
\end{eqnarray}
and $J\{a,\ b\}\equiv a_xb_y-b_xa_y$, as a starting point to
derive two component KdV equations by using the multiple scale
approach and the long wave approximation.

In \eqref{eq1}--\eqref{q2}, $F$ is the weak coupling constant
among two layer fluids and $\beta=\beta_0(L^2/U)$,
$\beta_0=(2\omega_0/a_0)\cos\phi_0$, in which $a_0$ is the earth's
radius, $\omega_0$ is the angular frequency of the earth's
rotation and $\phi_0$ is the latitude, $U$ is the characteristic
 velocity scales. The derivation of the dimensionless equation
 \eqref{eq1} and \eqref{eq2} is based on the characteristic horizontal length scale
$L=10^6m$ and the characteristic horizontal velocity scale
$U=10^{-1}m/s$ \cite{Rossby}.

More especially, if we take $\beta=0$, the system \eqref{q1} and
\eqref{q2} is reduced to the usual coupled Euler equation which is
suitable to be used to describe two-layer inviscid fluid. In other
words, all the results obtained in this paper are valid for
general two-layer inviscid fluid.

Under the long wave approximation in the $x$ direction, in order
to derive the KdV type equations, the stream functions $\psi_1$
and $\psi_2$ should have the form
\begin{eqnarray}
&&\psi_i=\phi_{i0}(y)+\phi_i(\epsilon (x-c_0t),\ y,\
\epsilon^3t)\equiv \phi_{i0}+\phi_i(X,\ y,\ T)\equiv
\phi_{i0}+\phi_i, \ i=1,\ 2,\label{psii}
\end{eqnarray}
where $\epsilon$ is a small parameter. It is reasonably considered
that the parameters $F$ and $\beta$ are in order $\epsilon$ and
$\epsilon^2$ respectively that means the coupling among two layers
is week and the effect of the rotation of the earth is much
smaller. Thus we set
\begin{eqnarray}
&&F=F_0\epsilon ,\qquad \beta=\beta_0 \epsilon^2.\label{F0}
\end{eqnarray}
Now, we expand the shift stream functions $\phi_i,\ i=1,\ 2$ as
\begin{eqnarray}
&&\phi_1=\epsilon \phi_{11}(X,\ y,\ T)+\epsilon^2 \phi_{12}(X,\
y,\ T)+\epsilon^3
\phi_{13}(X,\ y,\ T)+O(\epsilon^4),\label{phi1}\\
&&\phi_2=\epsilon \phi_{21}(X,\ y,\ T)+\epsilon^2 \phi_{22}(X,\
y,\ T)+\epsilon^3 \phi_{23}(X,\ y,\ T)+O(\epsilon^4).\label{phi2}
\end{eqnarray}
Substituting \eqref{psii}--\eqref{phi2} to
\eqref{eq1}--\eqref{eq2} yields
\begin{eqnarray}
&&\left[(\phi_{10y}-c)\partial_{yy}-\phi_{10yyy}\right]\phi_{11X}\epsilon^2
+\left\{\left[(\phi_{10y}-c)\partial_{yy}-\phi_{10yyy}\right]\phi_{12X}
+F_0(\phi_{10y}-c)\phi_{21X}\right.\nonumber\\
&&\left.
+[F_0(c_0-\phi_{20y})+\phi_{11yyy}]\phi_{11X}-\phi_{11y}\phi_{11yyX}\right\}\epsilon^3
+\left\{\left[(\phi_{10y}-c)\partial_{yyX}-\phi_{10yyy}\partial_X\right]\phi_{13}\right.\nonumber\\
&&\left.-\phi_{12y}\phi_{11yyX}-\phi_{11y}\phi_{12yyX}
+(\phi_{10y}-c_0)(F_0\phi_{22}+\phi_{11XX})_X+\phi_{11yyT}
-F_0\phi_{21X}\phi_{11y}\right.\nonumber\\
&&\left.+[\phi_{12yyy} +F_0\phi_{21y}+\beta_0]\phi_{11X}
+[F_0(c_0-\phi_{20y})
+\phi_{11yyy}]\phi_{12X}\right\}\epsilon^4+O(\epsilon^5)=0,\label{psi11}
\end{eqnarray}
and
\begin{eqnarray}
&&\left[(\phi_{20y}-c)\partial_{yy}-v_{0yyy}\right]\phi_{21X}\epsilon^2
+\left\{\left[(\phi_{20y}-c)\partial_{yy}-\phi_{20yyy}\right]\phi_{22X}
+F_0(\phi_{20y}-c)\phi_{11X}\right.\nonumber\\
&&\left.
+[F_0(c_0-\phi_{10y})+\phi_{21yyy}]\phi_{21X}-\phi_{21y}\phi_{21yyX}\right\}\epsilon^3
+\left\{\left[(\phi_{20y}-c)\partial_{yyX}-\phi_{20yyy}\partial_X\right]\phi_{13}\right.\nonumber\\
&&\left. -\phi_{22y}\phi_{21yyX}-\phi_{21y}\phi_{22yyX}
+(\phi_{20y}-c_0)(F_0\phi_{12}+\phi_{21XX})_X+\phi_{21yyT}
-F_0\phi_{11X}\phi_{21y}\right.\nonumber\\
&&\left.+[\phi_{22yyy} +F_0\phi_{11y}+\beta_0]\phi_{21X}
+[F_0(c_0-\phi_{10y})+\phi_{21yyy}]\phi_{22X}\right\}\epsilon^4+O(\epsilon^5)=0.\label{psi21}
\end{eqnarray}

Vanishing the $\epsilon^2$ terms of  \eqref{psi11} and
\eqref{psi21}, we have a special solution
\begin{eqnarray}
&&\phi_{11}=A_1(X,\ T)B_1(y)\equiv A_1B_1,\label{phi11}\\
&&\phi_{21}=A_2(X,\ T)B_2(y)\equiv A_2B_2,\label{phi21}
\end{eqnarray}
with $B_1$ and $B_2$ linked to $\phi_{10}$ and $\phi_{20}$ by
\begin{eqnarray}
&&U_{0yy}B_1-B_{1y}\phi_{10y}+C_1=0,\quad
\phi_{10}=U_0+c_0y,\label{u0yy}
\end{eqnarray}
and
\begin{eqnarray}
&&V_{0yy}B_2-B_{2y}\phi_{20y}+C_2=0,\quad
\phi_{20}=V_0+c_0y,\label{v0yy}
\end{eqnarray}
respectively with arbitrary constants $C_1$ and $C_2$.

By using the relations \eqref{phi11}--\eqref{v0yy}, the equations
obtained by vanishing the $\epsilon^3$ orders of  \eqref{psi11}
and \eqref{psi21} and integrating once with respect to $X$ become
\begin{eqnarray}
&&2\phi_{10y}\left(B_1\partial_{yy}-B_{1yy}\right)\phi_{12}+B_1\left[b_{11}A_1^2
-2F_0(B_1\phi_{20y}A_1+B_2\phi_{10y}A_2)\right]=0,\label{order31}\\
&&2\phi_{20y}\left(B_2\partial_{yy}-B_{2yy}\right)\phi_{22}+B_2\left[b_{21}A_2^2
-2F_0(B_2\phi_{10y}A_2-B_1\phi_{20y}A_1)\right]=0,\label{order32}
\end{eqnarray}
where the integrating functions have been dropped away and
\begin{eqnarray}
b_{11}\equiv B_1B_{1yyy}-B_{1y}B_{1yy},\ \quad b_{21}\equiv
B_2B_{2yyy}-B_{2y}B_{2yy}.
\end{eqnarray}

It is readily verified that
\begin{eqnarray}
\phi_{12}=(B_3A_1^2+B_4A_2+B_0A_1)B_1,\quad
\phi_{22}=(B_5A_2^2+B_6A_1+B_7A_2)B_2,\label{phi1222}
\end{eqnarray}
with $B_0,\ B_3,\ B_4,\ B_5,\ B_6$ and $B_7$ being functions of
$y$ and determined by
\begin{eqnarray}
&&B_{3y}=\frac{b_3}{B_1^2},\quad b_{3y}=-\frac{B_1b_{11}}{f_1},\quad B_{5y}=\frac{b_5}{B_2^2}
,\quad b_{5y}=-\frac{B_2b_{21}}{g_1},\label{b21}\\
&&B_{4y}=\frac{b_4}{B_1^2},\quad b_{4y}=-F_0B_2B_{1},\quad B_{6y}=\frac{b_6}{B_2^2},\quad b_{6y}=-F_0B_2B_{1},\ \label{b46}\\
&&B_{0y}=\frac{b_0}{B_1^2},\quad
b_{0y}=F_0B_1^2\frac{g_{1}}{f_1},\quad
B_{7y}=\frac{b_7}{B_2^2},\quad b_{7y}=F_0B_2^2\frac{f_{1}}{g_1}, \label{B07}\\
&& f_1=U_{0y},\quad g_1=V_{0y},
\end{eqnarray}
solves the third order equations \eqref{order31} and
\eqref{order32}.

Because of \eqref{phi11}, \eqref{phi21} and \eqref{phi1222}, the
fourth order of \eqref{psi11} and \eqref{psi21} become
\begin{eqnarray}
&&f_1\left(\partial_{yy}-B_1^{-1}B_{1yy}\right)\phi_{13X}+
B_{1yy}A_{1X T}+f_1B_1A_{1XXX}+F_0(g_1B_1B_4-f_1B_2B_7)A_{2X}\nonumber\\
&&\quad +2f_1F_0B_2B_5A_2A_{2X}-(\beta_0B_1-F_0g_1B_0B_1+F_0f_1B_2B_6)A_{1X}+B_4b_{11}(A_1A_{2})_{X}\nonumber\\
&&\quad +\left[2b_{11}B_0-2F_0g_1B_1B_3+
\frac{F_0g_1B_1}{f_1B_2}\left(\frac{c_1B_2}{f_1}-\frac{d_1B_1}{g_1}
-B_2B_{1y}+B_1B_{2y}\right)\right]A_1A_{1X}
\nonumber\\
&&\quad
+\frac1{2f_1^2}\left[b_{11}(6B_3f_1^2+3f_1B_{1y}-c_1)-f_1B_1b_{11y}\right]A_1^2A_{1X}=0,
\label{4th1}
\end{eqnarray}
and
\begin{eqnarray}
&&g_1\left(\partial_{yy}-B_2^{-1}B_{2yy}\right)\phi_{23X}
+B_{2yy}A_{2X T}+g_1B_2A_{2XXX}+F_0(g_1B_1B_0-f_1B_2B_6)A_{1X}\nonumber\\
&&\quad +2g_1F_0B_1B_3A_1A_{1X}-(\beta_0B_2+F_0f_1B_7B_2-F_0g_1B_1B_4)A_{2X}+B_6b_{21}(A_{1}A_2)_{X}\nonumber\\
&&\quad
+\left[2b_{21}B_7-2F_0f_1B_2B_5+\frac{F_0f_1B_2}{g_1B_1}\left(\frac{d_1B_1}{g_1}-\frac{c_1B_2}{f_1}
-B_1B_{2y}+B_2B_{1y}\right)\right]A_2A_{2X}
\nonumber\\
&&\quad
+\frac1{2g_1^2}\left[b_{21}(6B_5g_1^2+3g_1B_{2y}-d_1)-g_1B_2b_{21y}\right]A_2^2A_{2X}=0.\label{4th2}
\end{eqnarray}

In the usual studies to solve \eqref{4th1} and \eqref{4th2} type
equations, especially in the atmospheric and ocean dynamics, one
would take $\phi_{13}$ and $\phi_{23}$ as zero. However, whence
$\phi_{13}$ and $\phi_{23}$ are taken as zero, there may be a
non-consistence problem because $A_1$ and $A_2$ are only the
functions of $X$ and $T$ while the coefficients of \eqref{4th1}
and \eqref{4th2} are explicitly $y$-dependent. In generally,
equations \eqref{4th1} and \eqref{4th2} are not consistent except
that all the $y$-dependent coefficients are proportional each
other up to constant level. The detailed analysis of the equations
\eqref{4th1} and \eqref{4th2} with $\phi_{13}=\phi_{23}=0$ tells
that it is impossible to select the functions $B_0,\ B_1,\ ...,\
B_7,\ U_0$ and $V_0$ which are proportional each other and satisfy
the equations \eqref{u0yy}, \eqref{v0yy},\
\eqref{b21}--\eqref{B07} at the same time. To avoid this kind of
inconsistency in the traditional literature, an unreasonable and
unclear procedure is usually made, taking a $y$ average by
integrating the inconsistent equations with respect to the
variable $y$ from $y_1$ to $y_2$.

Nevertheless, it is possible to get some consistent and
significant solutions from \eqref{4th1} and \eqref{4th2} by taking
nonzero $\phi_{13}$ and $\phi_{23}$. In this paper, we only give
out a possible selection of $\phi_{13}$ and $\phi_{23}$ to derive
coupled KdV type equations for the quantities $A_1$ and $A_2$.

It is straightforward to verify that if we take
\begin{eqnarray}
\phi_{13}=r_1\int A_{1X}A_2{\rm d X}
+r_2A_1^3+r_3A_1^2+r_4A_1+r_5A_1A_2+r_6A_2^2+r_7A_2+r_8A_{1XX},
\label{p13}
\end{eqnarray}
\begin{eqnarray}
\phi_{23}=s_1\int A_{1X}A_2{\rm d X}
+s_2A_2^3+s_3A_2^2+s_4A_2+s_5A_1A_2+s_6A_1^2+s_7A_1+s_8A_{2XX},
\label{p23}
\end{eqnarray}
with
\begin{eqnarray}
&&r_i=B_1\int^y\frac1{B_1(y'')^{2}}\int^{y''}R_i(y'){\rm d}y'{\rm
d}y'',\nonumber\\
&&s_i=B_2\int^y\frac1{B_2(y'')^{2}}\int^{y''}S_i(y'){\rm d}y'{\rm
d}y'',\
i=1,\ 2,\ ...,\ 8, \nonumber\\
&&R_1=-\frac{\alpha_1B_1B_{1yy}}{f_1}, \
R_2=\frac{B_1}{6f_1^{3}}[B_1f_1b_{11y}+b_{11}(c_1-3f_1B_{1y}-6B_3f_1^3)],\nonumber\\
&&R_3=\frac{B_1^2F_0g_1}{2f_1}\left(2B_3+\frac{B_2B_{1y}-B_1B_{2y}}{B_2f_1}-\frac{c_1}{f_1^2}\right)
-\frac{B_1}{f_1}(\alpha_5B_{1yy}+b_{11}B_0)+\frac{F_0d_1B_1^3}{f_1^2B_2},\nonumber\\
&&R_4=-\frac{B_1}{f_1}(\beta_0B_1-F_0B_0B_1g_1+F_0f_1B_2B_6),\
R_5=-\frac{B_1}{f_1}(\alpha_3B_{1yy}+b_{11}B_4),\nonumber\\
&&R_6=-\frac{B_1}{f_1}[(\alpha_2-\alpha_5)B_{1yy}+F_0f_1B_{2}B_5],\
R_7=\frac{F_0B_1}{f_1}(g_1B_1B_4-f_1B_2B_7),\nonumber\\
&&R_8=-\frac{B_1}{f_1}(\alpha_4B_{1yy}+f_1B_1),\ S_8=-\frac{B_2}{g_1}(\delta_4B_{2yy}+g_1B_2),\nonumber\\
&&S_1=\frac{\delta_1B_2B_{2yy}}{g_1}, \
S_2=\frac{B_2}{6g_1^{3}}[B_2g_1b_{21y}+b_{21}(d_1-3g_1B_{2y}-6B_5g_1^3)],\nonumber\\
&&S_3=\frac{B_2^2F_0f_1}{2g_1}\left(2B_5+\frac{B_1B_{2y}-B_2B_{1y}}{g_1B_1}-\frac{d_1}{g_1^2}\right)
+\frac{c_1F_0B_2^3}{2g_1^2B_1}
+\frac{B_2}{g_1}(\delta_5B_{2yy}-b_{21}B_7),\nonumber\\
&&S_4=-\frac{B_2}{g_1}(\beta_0B_2-F_0B_7B_2f_1+F_0g_1B_1B_4),\
S_5=\frac{B_2}{g_1}(\delta_3B_{2yy}-b_{21}B_6),\nonumber\\
&&S_6=\frac{B_2}{g_1}[(\delta_2-\delta_5)B_{1yy}-F_0g_1B_{1}B_3],\
S_7=\frac{F_0B_2}{g_1}(g_1B_1B_0-f_1B_2B_6)\nonumber
\end{eqnarray}
for arbitrary $B_1$ and $B_2$, then $A_1$ and $A_2$ satisfy the
following coupled KdV system
\begin{eqnarray}
&&A_{1T}+\alpha_1A_2A_{1X}+(\alpha_2A_2^2+\alpha_3A_1A_2+\alpha_4A_{1XX}+\alpha_5A_1^2)_X=0,\label{ckdv1}\\
&&A_{2T}+\delta_1A_2A_{1X}+(\delta_2A_1^2+\delta_3A_1A_2+\delta_4A_{2XX}+\delta_5A_2^2)_X=0,\label{ckdv2}
\end{eqnarray}
where ten constants $\{\alpha_i,\ \delta_i,\ i=1,\ 2,\ 3,\ 4,\
5\}$ are arbitrary.

Now the important question is how to get some types of exact
solutions of the coupled KdV system. Before giving out some
concrete solutions, we try to make a Painlev\'e classification at
first. That means we are going to give some constraints on the
parameters $\{\alpha_i,\ \delta_i,\ i=1,\ 2,\ 3,\ 4,\ 5\}$ such
that the solutions of the model are single valued with respect to
an arbitrary singular manifold.

\section{Painlev\'e classification of the coupled KdV system}

The Painlev\'e test is one of the best way to study  nonlinear
systems. In this section, we take a standard Painlev\'e test by
using the Kruskal's simplification for the coupled KdV system.

To pass the Painlev\'e test, four steps are required. The leading
order analysis, the resonances determination, the test of the
primary branch and the test of the secondary branches.

The leading order analysis for the coupled KdV system
\eqref{ckdv1} and \eqref{ckdv2} around the arbitrary manifold
$\phi$ shows us that there are two
possible cases:\\
Case 1.
\begin{eqnarray}
A_1\sim \frac{u_0}{\phi^2},\quad A_2\sim
\frac{v_0}{\phi^2}.\label{lead1}
\end{eqnarray}
In this case the parameters $\{\alpha_i,\ \delta_i\}$ and $\{u_0,\
v_0\}$ are linked by
\begin{eqnarray}
&&2\alpha_5u_0^2+2\alpha_2v_0^2+(2\alpha_3+\alpha_1)u_0v_0+12\alpha_4u_0=0,\nonumber\\
&&2\delta_5v_0^2+2\delta_2u_0^2+(2\delta_3+\delta_1)u_0v_0+12\delta_4v_0=0.\
\label{lead1u0}
\end{eqnarray}
 Case 2.
\begin{eqnarray}
A_1\sim \frac{u_0}{\phi^2},\quad A_2\sim
\frac{v_0}{\phi}\label{lead21}
\end{eqnarray}
or equivalently
$$A_1\sim \frac{u_0}{\phi},\quad A_2\sim \frac{v_0}{\phi^2}$$
which will be not considered since the exchange symmetry $\{A_1,\
A_2,\ \alpha_i,\ \delta_i\}\leftrightarrow\{A_2,\ A_1,\ $ $
\delta_i,\ \alpha_i\}$ for the coupled KdV system \eqref{ckdv1}
and \eqref{ckdv2}.

The case \eqref{lead21} appears only for
\begin{eqnarray}
\delta_2=0,\
\delta_4=\frac{\alpha_4}{\alpha_5}(2\delta_1+3\delta_3),\
u_0=-6\frac{\alpha_4}{\alpha_5}.\label{lead21u0}
\end{eqnarray}

The resonance analysis for the first case \eqref{lead1} shows us
that the resonant points are located at
\begin{eqnarray}
-1,\ 4,\ 6,\ j_1,\ j_2,\ j_3=9-j_1-j_2,\label{res1}
\end{eqnarray}
where $j_1,\ j_2$ and $j_3$ are three roots of
\begin{eqnarray}
u_0v_0\delta_4\alpha_4(j-9)j^2+[(14\delta_4 v_0 u_0-u_0^2 (v_0
\delta_1+\delta_3 v_0+2 \delta_2 u_0)) \alpha_4
-v_0^2 \delta_4 (2 \alpha_2 v_0+u_0 \alpha_3)]j\nonumber\\
\quad +(24 \delta_4 v_0 u_0+2 u_0^2 (4 \delta_2 u_0+v_0 \delta_1+2
\delta_3 v_0)) \alpha_4+2 v_0^2 \delta_4 (u_0 \alpha_1+2 u_0
\alpha_3+4 \alpha_2 v_0)=0\label{res11}
\end{eqnarray}
for the variable $j$. Apart from the equivalent decoupled case for
both $A_1$ and $A_2$ satisfy completely decoupled KdV equations,
the positive integer conditions for the resonant points lead to
the following only nonequivalent subcases (i) $j_1=j_2=0,\ j_3=9$,
(ii)$j_1=0,\ j_2=1,\ j_3=8$, (iii)$j_1=0,\ j_2=2,\ j_3=7$,
(iv)$j_1=0,\ j_2=3,\ j_3=6$, (v)$j_1=0,\ j_2=4,\ j_3=5$,
(vi)$j_1=j_2=1,\ j_3=7$, (vii)$j_1=1,\ j_2=2,\ j_3=6$,
(viii)$j_1=1,\ j_2=3,\ j_3=5$, (ix)$j_1=j_2=2,\ j_3=5$,
(x)$j_1=2,\ j_2=3,\ j_3=4$.

The resonance analysis for the second case \eqref{lead21} gives
that the resonances will appear at
\begin{eqnarray}
-1,\ 0,\ 4,\ 6,\ j_1,\ j_2=6-j_1,\label{res2}
\end{eqnarray}
where $j_1$ and $j_2$ are two solutions of
\begin{eqnarray}
j(2\delta_1+3\delta_3)(j-6)+27\delta_3+22\delta_1=0\label{res21}
\end{eqnarray}
for the variable $j$. It is clear that the positive integer
conditions for the resonance points lead to the following four
nonequivalent subcases (a) $j_1=0,\ j_2=6$, (b)$j_1=1,\ j_2=5$,
(c)$j_1=2,\ j_2=4$, and (d)$j_1=3,\ j_2=3$.

To check all the resonances for the subcases (i)--(x) and (a)--(d)
yields the possible Painlev\'e integrable models under some
conditions for the model parameters $\alpha_i$ and $\delta_i$.
Here we
 just list the final nonequivalent models but omit the detail and tedious
analysis because the procedures are standard. The result shows us
that there are only six Painlev\'e integrable subcases of the
coupled KdV system \eqref{ckdv1} and \eqref{ckdv2}.\\
{\bf P-integrable model 1.}
\begin{eqnarray}
&&A_{1T}+\left(A_{1XX}+\frac12(c_2-c_1-c_1c_2)A_1^2
+c_1A_1A_2-\frac12A_2^2\right)_X=0,\nonumber\\
&&A_{2T}+\left(A_{2XX}+\frac12(c_1-c_2-1)A_2^2
+c_2A_1A_2-\frac12c_1c_2A_1^2\right)_X=0,\label{kdv1}
\end{eqnarray}
where $c_1$ and $c_2$ are arbitrary constants. For this type of
coupled KdV system \eqref{kdv1}, there are three branches with the
resonances $\{-1,\ 2,\ 3,\ 4,\ 4,\ 6\}$, $\{-1,\ 2,\ 3,\ 4,\ 4,\
6\}$ and $\{-1,\ -1,\ 4,\ 4,\ 6,\ 6\}$
respectively and all the resonance conditions satisfied.\\
{\bf P-integrable model 2.}
\begin{eqnarray}
A_{1T}+[A_{1XX}-(c+3)(c+6)A_1^2-{c^2}A_2^2]_X+2c[(c+6)A_{1X}A_2+(c+3)A_1A_{2X}]=0,\nonumber\\
A_{2T}+[A_{2XX}-c(c-3)A_2^2-(c+3)^2A_1^2]_X+2(c+3)[cA_2A_{1X}+(c-3)A_1A_{2X}]=0,\label{kdv2}
\end{eqnarray}
where $c$ is an arbitrary constant. For the model system
\eqref{kdv2} there is only one branch with the resonances located
at $\{-1,\ 1,\ 2,\ 4,\ 6,\ 6\}$ and all the
resonance conditions satisfied identically.\\
{\bf P-integrable model 3.}
\begin{eqnarray}
&&A_{1T}+(A_{1XX}+A_1^2+A_1A_2)_X=0,\nonumber\\
&&A_{2T}+(A_{2XX}+A_2^2+A_1A_2)_X=0.\label{kdv3}
\end{eqnarray}
In this case, the resonance points are $\{-1,\ 0,\ 4,\ 4,\ 5,\ 6\}$.\\
{\bf P-integrable model 4.}
\begin{eqnarray}
&&A_{1T}+[A_{1XX}+(A_1+A_2)^2]_X=0,\nonumber\\
&&A_{2T}+[A_{2XX}+(A_1+A_2)^2]_X=0.\label{kdv4}
\end{eqnarray}
This case is corresponding to the resonances are located at $\{-1,\ 2,\ 3,\ 4,\ 4,\ 6\}$.\\
{\bf P-integrable model 5.}
\begin{eqnarray}
&&A_{1T}+(A_{1XX}+A_1^2)_X+2A_2A_{1X}=0,\nonumber\\
&&A_{2T}+(A_{2XX}+A_2^2)_X+2A_1A_{2X}=0.\label{kdv5}
\end{eqnarray}
This case is corresponding to the resonances are located at $\{-1,\ 0,\ 2,\ 4,\ 6,\ 7\}$.\\
{\bf P-integrable model 6.}
\begin{eqnarray}
&&A_{1T}+A_{1XXX}+(A_1+A_2)(3A_{1}+A_{2})_X=0,\nonumber\\
&&A_{2T}+A_{2XXX}+(A_1+A_2)(3A_{2}+A_1)_{X}=0.\label{kdv6}
\end{eqnarray}
This case is corresponding to the resonances are located at
$\{-1,\ 1,\ 2,\ 4,\ 6,\ 6\}$.

\section{Exact solutions}
In this section, we study some types of exact solutions for the
general couple KdV system \eqref{ckdv1} and \eqref{ckdv2} and some
special types of P-integrable models.

\subsection{Travelling periodic and solitary wave solutions of the
general coupled KdV system \eqref{ckdv1} and \eqref{ckdv2}} In
\cite{JMP1989Lou}, it is pointed out that some special types of
exact solutions including travelling wave solutions of various
nonlinear systems can be obtained via the deformation and mapping
approach from the solutions of the cubic nonlinear Klein-Gordon
equation (or namely $\phi^4$ model). It is quite easy to see that
for some types of travelling wave solutions of the coupled KdV
system \eqref{ckdv1} and \eqref{ckdv2} can also be obtained by
some suitable deformation relations from the travelling wave
solution of the $\phi^4$ model.

For the travelling wave solution of the coupled KdV system
\eqref{ckdv1} and \eqref{ckdv2},
\begin{eqnarray}
A_1=A_1(\xi)\equiv A_1(kX-k c T), \ A_2=A_2(\xi),\label{trav}
\end{eqnarray}
we have
\begin{eqnarray}
\alpha_1A_{1\xi}A_2+(\alpha_2A_2^2+\alpha_3A_1A_2+\alpha_4k^2A_{1\xi\xi}+\alpha_5A_1^2-cA_1)_\xi=0,\\
\delta_1A_{1\xi}A_2+(\delta_2A_1^2+\delta_3A_1A_2+\delta_4k^2A_{2\xi\xi}+\delta_5A_2^2-cA_2)_\xi=0,
\label{Trkdv}
\end{eqnarray}
To map the travelling waves of the cubic nonlinear Klein-Gordon
equation to those of the coupled KdV system, one may use different
mapping relations such as the polynomial forms \cite{JMP1989Lou},
rational forms \cite{CTP1993Lou} and may be more complicated
and/or derive dependent forms \cite{PRE2002Chen}. However, here we
just give a simple polynomial deformation relation
\begin{eqnarray}
A_1=a_0+a_1\phi(\xi)+a_2\phi(\xi)^2,\quad
A_2=b_0+b_1\phi(\xi)+ba_2\phi(\xi)^2,\ \label{A12}
\end{eqnarray}
where $\phi(\xi)$ is a travelling wave solution of the cubic
nonlinear Klein-Gordon equation, i.e., $\phi$ satisfies
\begin{eqnarray}
\phi_\xi^2=\mu\phi^2+\frac12\lambda\phi^4+C.\label{phi4}
\end{eqnarray}

It is not very difficult to find that $\{A_1,\ A_2\}$ given by
\eqref{A12} with \eqref{phi4} is a solution of the coupled KdV
system \eqref{ckdv1} and \eqref{ckdv2} if and only if the eleven
solution parameters $a_0,\ a_1,\ a_2,\ b_0,\ b_1,\ b,\ \mu,\
\lambda,\ C,\ k,\ c$ and ten model parameters $\alpha_i,\
\delta_i,\ i=1,\ 2,\ ...,\ 8$ satisfy the following eight
constraints
\begin{eqnarray}
&&(2 \alpha_1+3 \alpha_3+6 \alpha_2 b) a_2 b_1+[(3 \alpha_3 b+6
\alpha_5+\alpha_1 b) a_2+3 k^2 \alpha_4 \lambda] a_1=0,\nonumber\\
&& (2 \alpha_3 b+4 \alpha_5) a_2 a_0+2 \alpha_2 b_1^2+(\alpha_1+2
\alpha_3) a_1 b_1+2 \alpha_5 a_1^2\nonumber\\
&&\quad  +[8 k^2 \alpha_4 \mu-2 c+(2
\alpha_3+4 \alpha_2 b+2 \alpha_1) b_0] a_2=0,\nonumber\\
&& (2 \alpha_5 a_1+b_1 \alpha_3) a_0+2 \alpha_2 b_0 b_1+[k^2
\alpha_4
\mu-c+(\alpha_1+\alpha_3) b_0] a_1=0,\nonumber\\
&& (b_1 \delta_3+2 \delta_2 a_1) a_0+(k^2 \delta_4 \mu+2 b_0
\delta_5-c)
b_1+(\delta_1+\delta_3) b_0 a_1=0,\nonumber\\
&& (4 \delta_5 b^2+2 \delta_1 b+4 \delta_2+4 \delta_3 b) a_2+12
k^2 \delta_4 b
\lambda=0,\nonumber\\
&&(4 \delta_2+2 \delta_3 b) a_2 a_0+2 \delta_5 b_1^2+(2
\delta_3+\delta_1) a_1 b_1+2 \delta_2 a_1^2 +[8 k^2 \delta_4 b
\mu\nonumber\\
&&\quad -2 c b+(4 \delta_5 b+2 \delta_1+2 \delta_3) b_0] a_2=0,\\
&&(4 \alpha_3 b+2 \alpha_1 b+4 \alpha_2 b^2+4 \alpha_5) a_2+12
k^2 \alpha_4  \lambda=0,\nonumber\\
&&[(6 \delta_5 b+2 \delta_1+3 \delta_3) a_2+3 k^2 \delta_4
\lambda] b_1+(3 \delta_3 b+\delta_1 b+6 \delta_2) a_2
a_1=0.\label{abcd}
\end{eqnarray}
It is obvious that the algebraic equation system \eqref{abcd}
possesses many kinds of solutions. Here we just write down a most
important and simplest solution when $\delta_4=\alpha_4$:
\begin{eqnarray}
a_0=a_1=b_0=b_1=0,\ c=4k^2\mu \alpha_4,\
a_2=-\frac{6k^2\lambda\alpha_4}{2\alpha_5+2b\alpha_3+b\alpha_1+2b^2\alpha_2},
\end{eqnarray}
while $b$ is linked the model parameters by a cubic algebraic
equation
\begin{eqnarray}
\delta_2+\left(\delta_3-\alpha_5+\frac12\delta_1\right)b
+\left(\delta_5-\alpha_3-\frac12\alpha_1\right)b-\alpha_2b^3=0.\label{rb}
\end{eqnarray}
More concretely, if we take $\phi(\xi)$ as the Jacobi elliptic
conoid function
$$\phi={\rm cn}(\xi,\ m)$$ which is a special solution of the $\phi^4$ model with the parameters
$$\mu = 2 m^2-1,\ \lambda =-2m^2 ,\ C = 1-m^2,$$
then we have a simple periodic wave solution for the coupled KdV
equation \eqref{ckdv1} and \eqref{ckdv2} with $\delta_4=\alpha_4$
\begin{eqnarray}
A_1=\frac{12k^2
m^2\alpha_4}{2\alpha_5+2b\alpha_3+b\alpha_1+2b^2\alpha_2}{\rm
cn}^2
\left(kX-4k^3(2m^2-1)\alpha_4T,\ m\right),\nonumber\\
A_2=\frac{12k^2
m^2\alpha_4b}{2\alpha_5+2b\alpha_3+b\alpha_1+2b^2\alpha_2}{\rm
cn}^2\left(kX-4k^3(2m^2-1)\alpha_4T,\ m\right),\label{ps}
\end{eqnarray}
where $b$ is a solution of \eqref{rb}. Furthermore, when $m=1$,
the periodic solution \eqref{ps} becomes a simple solitary wave
solution
\begin{eqnarray}
A_1=\frac{12k^2\alpha_4}{2\alpha_5+2b\alpha_3+b\alpha_1+2b^2\alpha_2}{\rm
sech}^2
\left(kX-4k^3\alpha_4T\right),\nonumber\\
A_2=\frac{12k^2\alpha_4b}{2\alpha_5+2b\alpha_3+b\alpha_1+2b^2\alpha_2}{\rm
sech}^2\left(kX-4k^3\alpha_4T\right).\label{ss}
\end{eqnarray}
\subsection{$\tau$-function solutions and Multi-soliton solutions of the coupled KdV system}

\subsubsection{The first type of $\tau$-function and multi-soliton solutions related to the KdV reductions.}
It is straightforward to verify that for the coupled KdV equation
system \eqref{ckdv1} and \eqref{ckdv2} with
\begin{eqnarray}
\delta_4=\alpha_4,\label{ad4}
\end{eqnarray}
one can find at least one type of multiple soliton solutions
because there is a simple KdV reduction
\begin{eqnarray}
A_{1T}+\alpha_4A_{1XXX}+(a\alpha_1+2\alpha_2a^2+2a\alpha_3+2\alpha_5)A_1A_{1X}=0,\
A_2=aA_1, \label{type1}
\end{eqnarray}
where $a$ is a solution of the algebraic cubic equation
\begin{eqnarray}
2a^3\alpha_2+(\alpha_1+2\alpha_3-2\delta_5)a^2+(2\alpha_5-\delta_1-2\delta_3)a-2\delta_2=0.\label{cubic}
\end{eqnarray}
Then the coupled KdV equation system \eqref{ckdv1} and
\eqref{ckdv2} with \eqref{ad4} possesses the following $\tau$
function and multiple soliton solutions
\begin{eqnarray}
A_1=\frac{A_2}a=\frac{12\alpha_4}{a\alpha_1+2\alpha_2a^2+2a\alpha_3+2\alpha_5}\left(\ln
\tau \right)_{XX},
\end{eqnarray}
where $\tau$ is just the usual $\tau$ function. For the
multi-soliton solution, the $\tau$ function reads
\begin{eqnarray}
&&\tau=1+\sum_{i=1}^N \pi_i+\sum_{i_1<i_2}^N
A_{i_1i_2}\pi_{i_1}\pi_{i_2}+\sum_{i_1<i_2<i_3}^N
A_{i_1i_2i_3}\pi_{i_1}\pi_{i_2}\pi_{i_3}+...+A_{i_1i_2...i_N}\pi_{i_1}\pi_{i_2}...\pi_{i_N},\nonumber
\\
&&\pi_i\equiv \exp(k_iX-\alpha_4k_i^3T),\ \quad
A_{i_1i_2...i_k}\equiv \prod_{i_a<i_b,a,b=1,2,...k}A_{i_ai_b}.
\label{phi}
\end{eqnarray}

It is interesting that there is only one parameter condition
\eqref{ad4} to get multiple soliton solution \eqref{type1} while
it has been proved that the model is non-Painlev\'e integrable. In
other words, the existence condition for multiple soliton
solutions is not a sufficient condition of the integrability.

Especially, because of there are three real solutions of
\eqref{cubic} for the special coupled KdV equation \eqref{kdv1},
we can obtain three types of multiple soliton solutions $\{u_1,\
v_1\}$, $\{u_2,\ v_2\}$ and $\{u_3,\ v_3\}$,
\begin{eqnarray}
v_1=u_1=\frac{12}{(c_1-1)(c_2-1)}(\ln \tau)_{XX},
\end{eqnarray}
\begin{eqnarray}
u_2=\frac{12}{(c_1-1)(c_1-c_2)}(\ln \tau)_{XX},\ v_2={c_1}u_2,
\end{eqnarray}
and
\begin{eqnarray}
u_3=\frac{12}{(c_2-1)(c_1-c_2)}(\ln \tau)_{XX},\ v_3={c_2}u_2
\end{eqnarray}
with $\tau$ being given by \eqref{phi}.

\subsubsection{The second type of $\tau$-function and multi-soliton solutions of the coupled KdV system.}
The multi-solitons of the coupled KdV system listed in the last
subsection are obtained from its special KdV reduction. In
\cite{Lou1999map} it has been found that even if for the
non-integrable cases, the coupled nonlinear system may have much
more abundant solitary wave structure. So we believe that for the
coupled KdV system there may be other types of multiple soliton
solutions.

For instance, if the model parameters have the following
conditions
\begin{eqnarray}
&&\alpha_1\alpha_2(\alpha_1\delta_3-\delta_1\alpha_3)\neq 0,\
\delta_4=\alpha_4,\ \nonumber\\
&&\delta_5 =
\frac12\alpha_3+\frac{\alpha_2\delta_1}{\alpha_1}-\frac{\alpha_1\delta_3}{2\delta_1},\
\alpha_5 =
-\frac{\delta_1\alpha_3}{2\alpha_1}+\frac12\delta_3+\frac{\alpha_1\delta_2}{\delta_1},\\
&&\alpha_3=-\frac{2\delta_2\alpha_1^2}{\delta_1^2}-\frac{\alpha_1(\delta_1+\delta_3)}{\delta_1}
-\frac{2\delta_1\alpha_2}{\alpha_1}, \label{a3}
\end{eqnarray}
then we have a new type of multiple soliton solution
\begin{eqnarray}
A_1=\frac{12\alpha_1\alpha_4}{\alpha_1\delta_3-\delta_1\alpha_3}\left(\ln
\tau \right)_{XX}+\frac{\alpha_1}{\delta_1}A_2, \
\end{eqnarray}
where $\tau$ is still the $\tau$ function of the usual KdV
equation and for the multi-soliton solution it is given by
\eqref{phi} while $A_2$ is linked to the $\tau$-function by a \em
linear \rm equation
\begin{eqnarray}
&&A_{2T}+\frac{12\alpha_1\alpha_4
(\delta_1\delta_3+2\delta_2\alpha_1)}{\delta_1\alpha_3-\alpha_1\delta_3}A_{2X}(\ln
\tau
)_{XX}+\frac{144\delta_2\alpha_1^2\alpha_4^2}{(\delta_1\alpha_3-\alpha_1\delta_3)^2}[(\ln
\tau )_{XX}^2]_X\nonumber\\
&&\quad
-\frac{12\alpha_1\alpha_4(\delta_1\delta_3+\delta_1^2+2\delta_2\alpha_1)}{\delta_1\alpha_3-\alpha_1\delta_3}A_2(\ln
\tau )_{XXX}+\alpha_4A_{2XXX}=0, \label{tau2}
\end{eqnarray}
If the third condition \eqref{a3} is not satisfied, then a
$nonlinear$ term
$$\left({\delta_1+\delta_3}+\frac{\delta_1\alpha_3}{\alpha_1}+\frac{2\delta_2\alpha_1}{\delta_1}
+\frac{2\delta_1^2\alpha_2}{\alpha_1^2}\right)A_2A_{2X} $$
 has to be added
to the left hand side of \eqref{tau2}.

Similarly, under the conditions
\begin{eqnarray}
&&\alpha_3\neq 0,\ \alpha_1=\delta_1=0,\ \delta_4=\alpha_4, \
\delta_5=\frac{\alpha_2(\delta_3-c_1)^2}{c_1\alpha_3^3}-\frac12,\label{a31} \\
&&\quad \alpha_2
=-\frac{\alpha_3^2(2c_1\delta_3+2\delta_2-c_1^2)}{2c_1(c_1-\delta_3)^2}\label{a32}
\end{eqnarray}
where the constant $c_1$ is determined by
\begin{eqnarray}
c_1=\alpha_5\pm \sqrt{\alpha_5^2-2\delta_2},
\end{eqnarray}
we have the following new type of multiple soliton solutions
\begin{eqnarray}
A_1=\frac{12\alpha_4}{c_1}\left(\ln \tau
\right)_{XX}+\frac{\alpha_3}{\delta_3-c_1}A_2, \
\end{eqnarray}
where $\tau$ is same as given by \eqref{phi} while $A_2$ is also
linked to the $\tau$ function by a \em linear \rm equation
\begin{eqnarray}
A_{2T}+\alpha_4A_{2XXX}+\frac{12\alpha_4
(c_1\delta_3+2\delta_2)}{c_1^2}[A_{2}(\ln \tau
)_{XX}]_X+\frac{144\delta_2(\delta_3-c_1)\alpha_4^2}{\alpha_3c_1^2}[(\ln
\tau )_{XX}^2]_X=0. \label{tau3}
\end{eqnarray}
In the same way, if the parameter condition \eqref{a32} is not
satisfied, then we have to add a nonlinear term
$$\left(\frac{2\alpha_2(\delta_3-c_1)}{\alpha_3}
+\frac{\alpha_3(2c_1\delta_3+2\delta_2-c_1^2)}{c_1(\delta_3-c_1)}\right)A_2A_{2X}$$
to the left hand side of \eqref{tau3}.

\subsubsection{The third type of $\tau$-function and multi-soliton solutions of the coupled KdV system.}

Actually, in additional to the above types of multiple soliton
solutions there may be other types of soliton solutions. Here is a
further simple example for the following more specifical model
\begin{eqnarray}
&&A_{1T}+aA_{1XXX}+bA_1A_{1X}+bcA_2A_{2X}=0,\nonumber\\
&&A_{2T}+aA_{2XXX}+bA_1A_{2X}+bA_2A_{1X}=0.\label{sckdv}
\end{eqnarray}
For this special model, its first type of multiple soliton
solutions has the form
\begin{eqnarray}
&&A_1= \frac{6a}{b}(\ln \tau)_{XX},\ A_2=\pm \frac{1}{\sqrt{c}}A_1
\end{eqnarray}
and the second type of multiple soliton solutions is given by
\begin{eqnarray}
&&A_1= \frac{12a}{b}(\ln \tau)_{XX}\pm \sqrt{c}A_2,\
\end{eqnarray}
while $A_2$ satisfies
\begin{eqnarray}
&&A_{2T}+aA_{2XXX}+12a[A_{2}(\ln \tau)_{XX}]_X\pm
2b\sqrt{c}A_2A_{2X}=0.
\end{eqnarray}

For the special coupled KdV system \eqref{sckdv}, there is the
following third type of multiple soliton solutions,
\begin{eqnarray}
&&A_1= \frac{6a}{b}[\ln (\tau_1^2+\tau_2^2)]_{XX},\label{tsolA1}\\
&&A_2=\pm \frac{12a}{b\sqrt{-c}}\left({\rm arctan}
\frac{\tau_2}{\tau_1}\right)_{XX},\label{tsolA2}
\end{eqnarray}
where $$\tau\equiv \tau_1+i\tau_2$$ is just the usual $\tau$
function of the KdV equation but with \em complex \rm parameters,
$\tau_1$ and $\tau_2$ are the real and imaginary parts of $\tau$
respectively.

\input epsf
\begin {figure}
\centering \epsfxsize=10cm\epsfysize=3cm\epsfbox{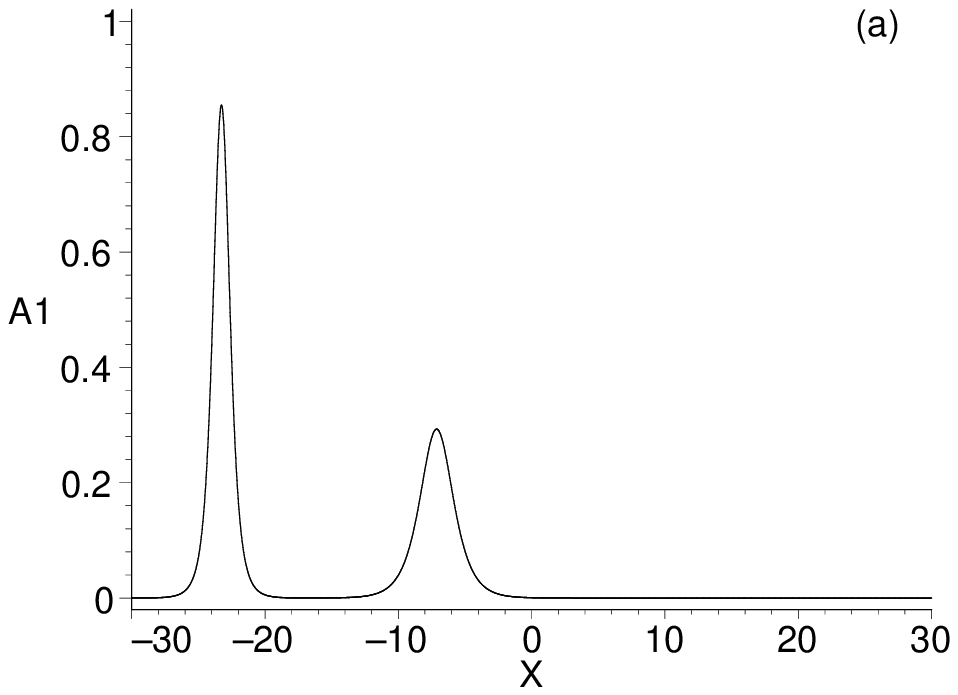}
\epsfxsize=10cm\epsfysize=3cm\epsfbox{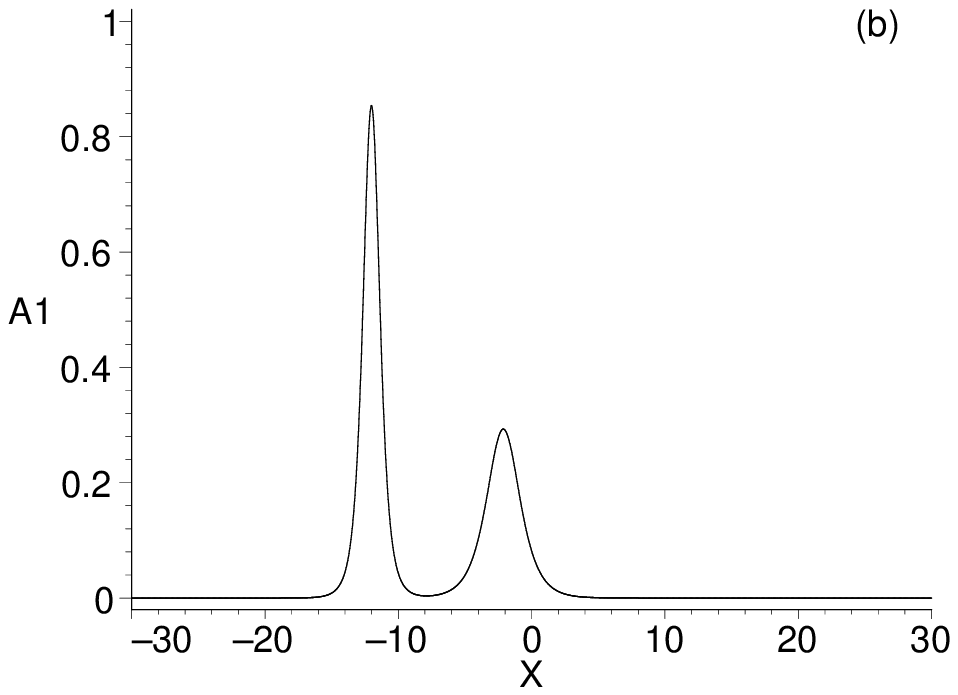}
\epsfxsize=10cm\epsfysize=3cm\epsfbox{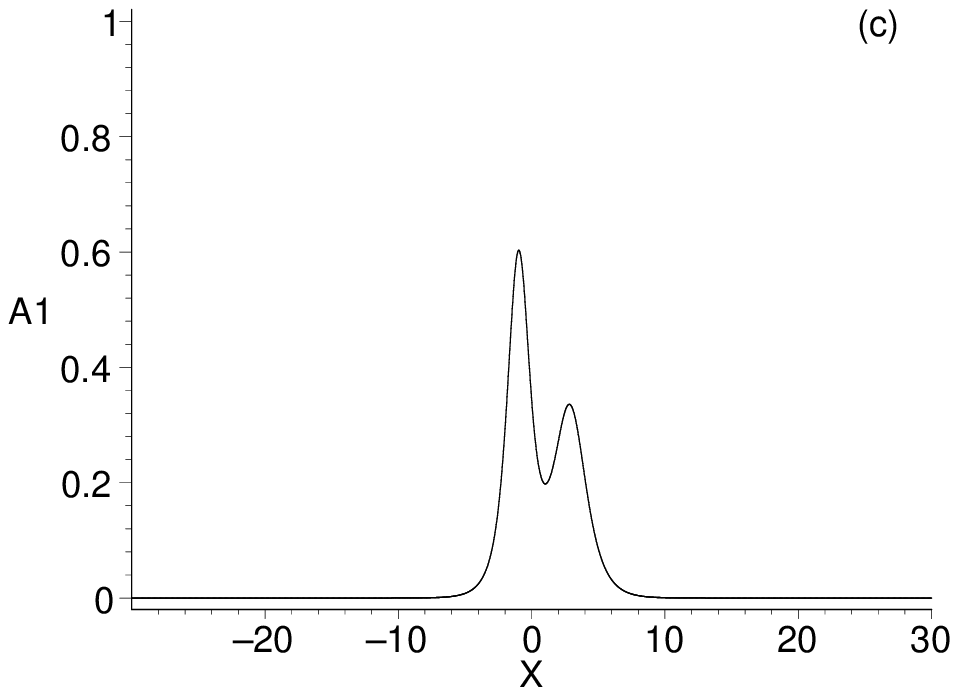}
\epsfxsize=10cm\epsfysize=3cm\epsfbox{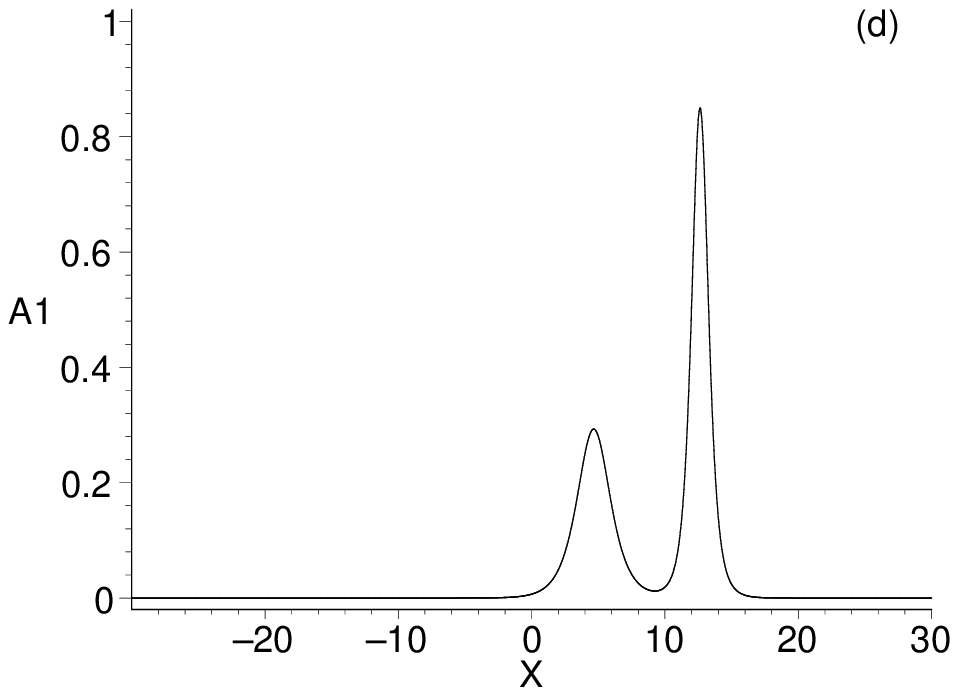}
\epsfxsize=10cm\epsfysize=3cm\epsfbox{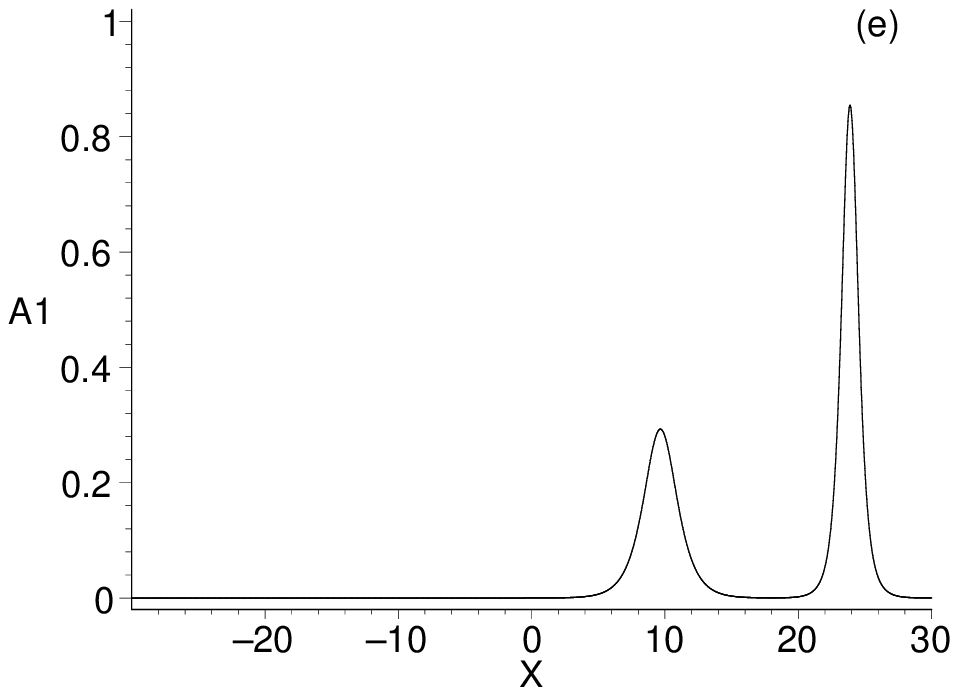}
 \caption{The interaction plots of two soliton solution for the field
 $A1\equiv A_1$ expressed by \eqref{tsolA1}
 and \eqref{2sol} at times (a) $T=-10$, (b) $T=-5$,
(c) $T=-0$, (d) $T=5$ and (e) $T=10$ respectively.}
\end{figure}

\input epsf
\begin {figure}
\centering \epsfxsize=10cm\epsfysize=3cm\epsfbox{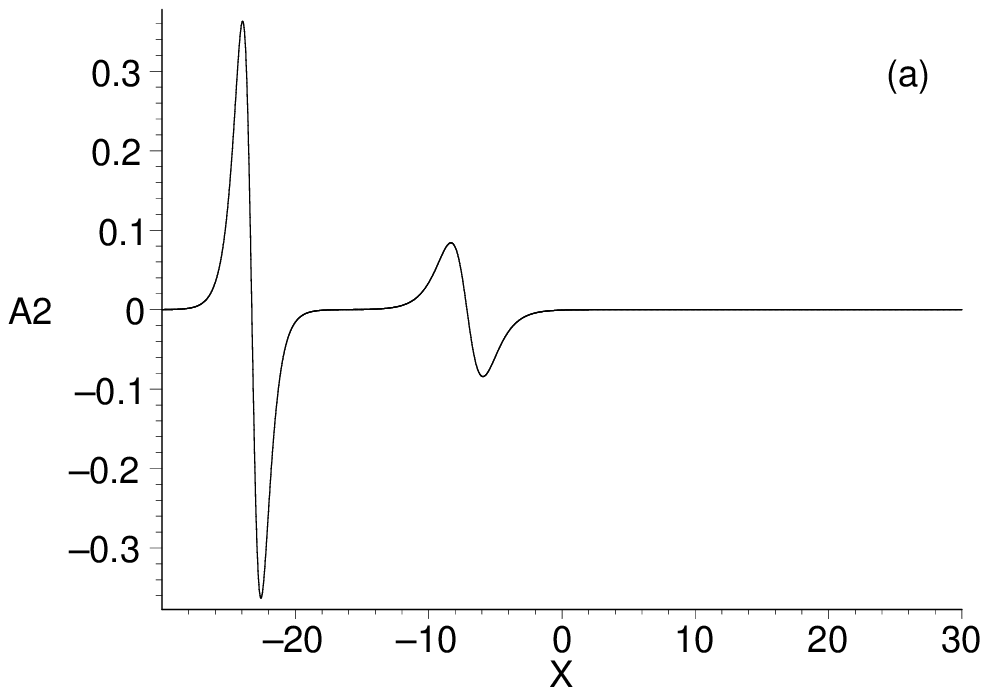}
\epsfxsize=10cm\epsfysize=3cm\epsfbox{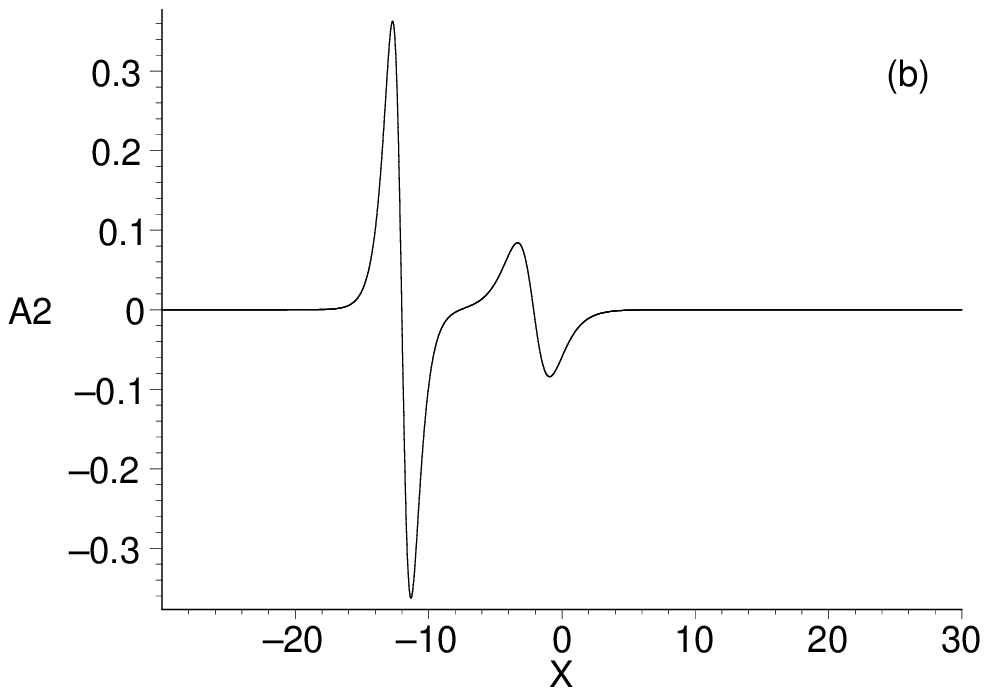}
\epsfxsize=10cm\epsfysize=3cm\epsfbox{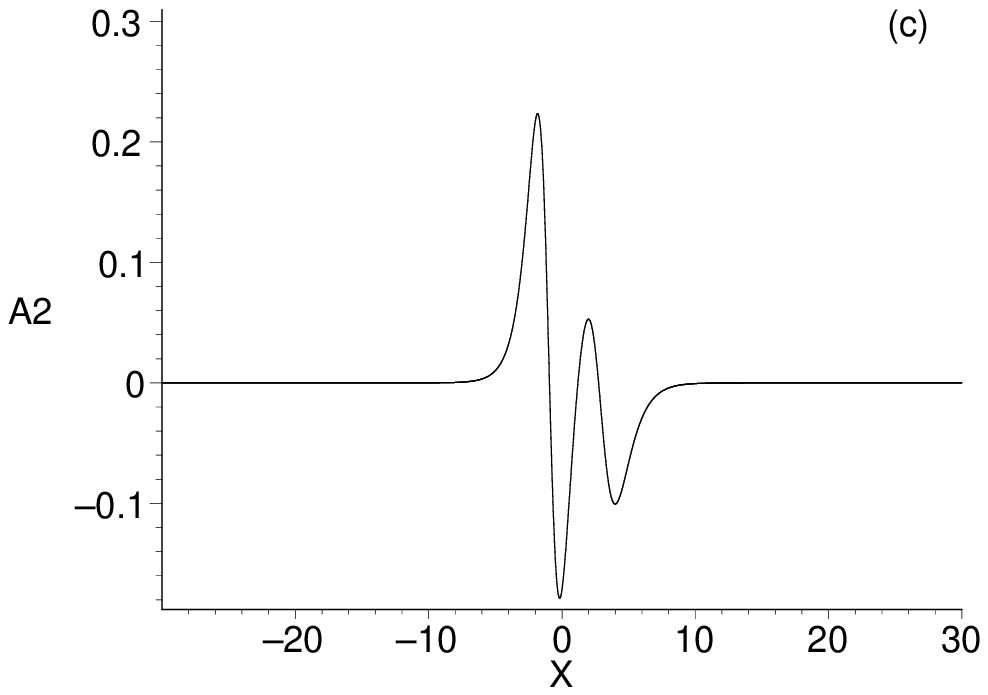}
\epsfxsize=10cm\epsfysize=3cm\epsfbox{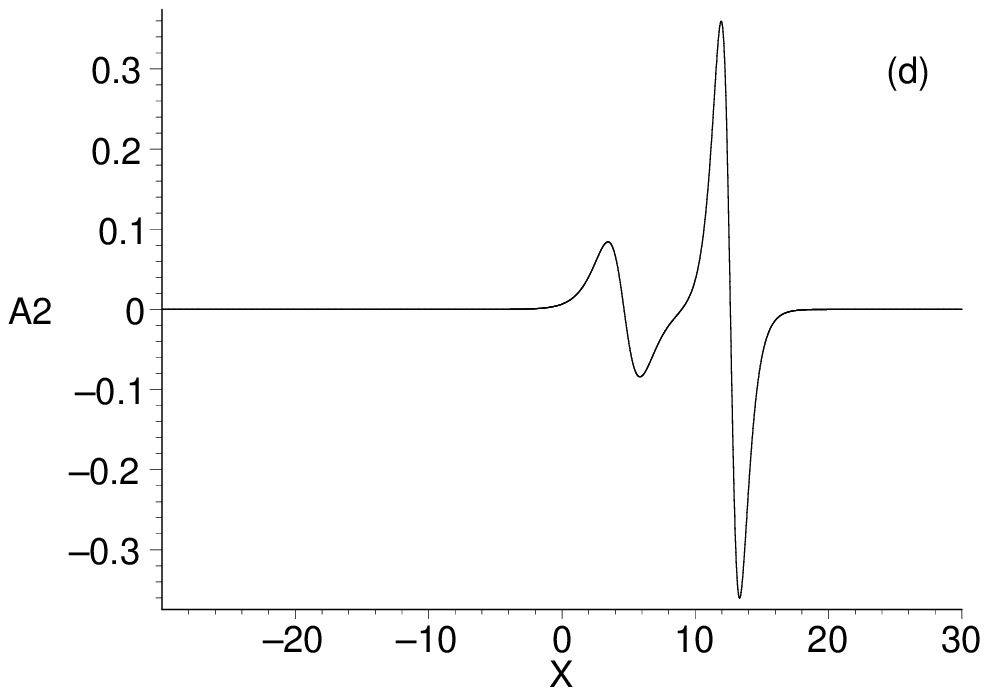}
\epsfxsize=10cm\epsfysize=3cm\epsfbox{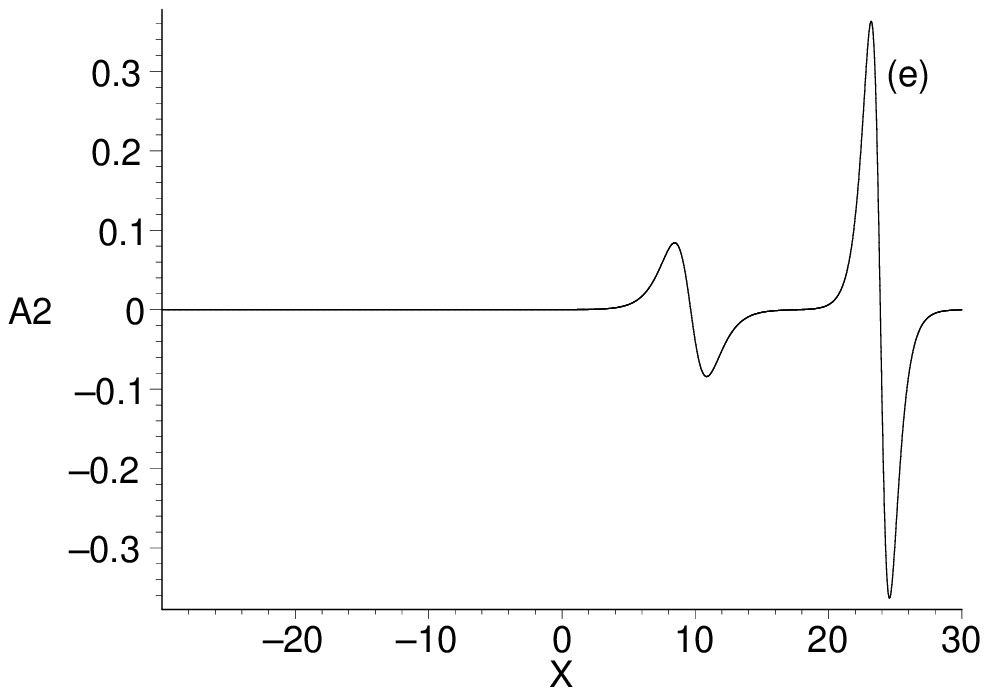}
 \caption{The interaction plots of the two-soliton solution for the field $ A2\equiv A_2$ expressed by \eqref{tsolA2}
 and \eqref{2sol} at times (a) $T=-10$, (b) $T=-5$, (c) $T=-0$, (d) $T=5$ and (e) $T=10$ respectively.}
\end{figure}

Fig. 1 and Fig. 2 are the special interaction plots of two soliton
solution for the coupled KdV system \eqref{sckdv} of the field
$A_1$ and $A_2$ expressed by \eqref{tsolA1} and \eqref{tsolA2}
respectively with
\begin{eqnarray}
&&\tau=1+(1+i)e^{k_1X-k_1^3T}+(1+3i)e^{k_2X-k_2^3T}
+\frac{(k_1-k_2)^2}{(k_1+k_2)^2}(4i-2)e^{(k_1+k_2)X-(k_1^3+k_2^3)T},\nonumber\\
&&k_1=1,\quad k_2=\frac32\label{2sol}
\end{eqnarray}
at times $T=-10,\ -5,\ 0,\ 5$ and $10$ respectively.

\section{Summary and discussions}
In summary, a general type of coupled KdV system is derived from
the coupled Euler equation system \eqref{eq1} and \eqref{eq2} with
($\beta\neq 0$) and/or without ($\beta\neq 0$) the consideration
of the earth rotation effects. In the derivation procedure, the
frequently used inconsistent $y$-average trick in the past
literature is removed.

The integrability of the derived KdV system is checked by means of
the well known Weiss-Tabor-Carnevale's Painlev\'e test procedure.
It is found that there are six types of Painlev\'e integrable
subcases for the derived coupled KdV system.

The deformation and mapping method are used to get some types of
travelling wave solutions including the conoidal periodic waves
and single solitary waves for the general derived coupled KdV
system with \eqref{ad4}.

It is found that the coupled nonlinear systems may possesses much
more abundant solution structures. This phenomena is found before
for the coupled non-integrable high-dimensional Klein-Gordon
equation \cite{Lou1999map}. In this paper we found that whence
some kinds of model parameter conditions are satisfied, then there
may be some different kinds of \em multiple \rm soliton solutions
and $\tau$ function solutions. On the other hand, for the coupled
KdV equation obtained here

The dynamics of atmospheric blockings has been one of the central
and important problem since they are the main representations of
the general circulation anormaly in the areas of mid-high
latitudes. Atmospheric blocking events have a strong influence on
regional weather and climate. The observations have shown
atmospheric blockings may locate in the mid-high latitudes,
usually over the ocean, as the dipole pattern which was first
discovered by Rex \cite{Rex}.

For one layer atmospheric model, the single soliton solution of
the constant coefficient KdV equation is responsible for the
dipole pattern of the atmospheric blockings. To explain the
blocking life cycle, one has to use the soliton solutions of the
variable coefficient KdV equation \cite{THL}. Using the similar
analysis as the single layer atmospheric model, the soliton
solution of the coupled KdV equation can also be used to explain
the blockings under the two layer atmospheric description. In the
same way, the soliton solutions of the constant coefficient
coupled KdV equation proposed here can not be used to describe the
blocking life cycle. To describe the blocking life cycle one has
to extended the coupled KdV system \eqref{eq1} and \eqref{eq2} to
variable coefficient case and the problem will be detailed studied
in the near future study.

The authors are indebt to discuss with Dr. F. Huang and Prof. Y.
Chen. The work was supported by the National Natural Science
Foundation of China (No. 90203001, No. 10475055 and No. 40305009).

\end{document}